\documentclass[prl,twocolumn,showpacs]{revtex4}% Physical Review B
\usepackage{graphicx,bm}% Include figure files
\input{epsf.sty}
\begin{document}
%\preprint{}

\title{Thermal conductivity evidence for $d_{x^2-y^2}$ pairing symmetry \\in the heavy-fermion CeIrIn$_5$ superconductor }

%{}%

\author{Y.\ Kasahara$^\mathrm{1}$,  T.\ Iwasawa$^\mathrm{1}$, Y.\ Shimizu$^\mathrm{1}$, H.\ Shishido$^\mathrm{1}$, T.\ Shibauchi$^\mathrm{1}$,  I.~Vekhter$^\mathrm{2}$, and Y.\ Matsuda$^\mathrm{1,3}$}
\affiliation{$^1$Department of Physics, Kyoto University, Kyoto 606-8502, Japan }%
\affiliation{$^2$Department of Physics and Astronomy, Louisiana State University, Baton Rouge, Louisiana, 70803, USA}%
\affiliation{$^3$Institute for Solid State Physics, University of Tokyo, Kashiwanoha, Kashiwa, Chiba 277-8581, Japan}%

%\date{\today}

\begin{abstract}

Quasi-two dimensional CeIrIn$_5$ contains two distinct domes with
different  heavy fermion superconducting states in its phase
diagram.  Here we pinned down the superconducting gap structure of
CeIrIn$_5$ in the second dome, located away from the
antiferromagnetic (AF) quantum critical point,   by the thermal
transport measurements in magnetic fields {\boldmath $H$} rotated
relative to the crystal axes.  Clear fourfold oscillation was
observed when {\boldmath $H$} is rotated within the $ab$-plane,
while no oscillation was observed within the $bc$-plane.  In sharp
contrast to previous reports, our results are most consistent with
$d_{x^2-y^2}$ symmetry, implying that two superconducting phases
have the same gap symmetry which appears to be mediated by AF spin
fluctuations.

\end{abstract}

\pacs{74.20.Rp,74.25.Dw,74.25.Fy,74.70.Tx}
%74.20.Rp   Pairing symmetries (other than s-wave)
%74.25.Dw   Superconductivity phase diagrams
%74.25.Fy   Transport properties (electric and thermal conductivity, thermoelectric effects, etc.)
%74.70.Tx   Heavy-fermion superconductors

\maketitle

Superconductivity in heavy-Fermion (HF) compounds continues to  be
a central focus of investigations into strongly correlated
electron systems.  The relationship between the magnetism and
unconventional superconductivity, whereby the gap function
$\Delta$({\boldmath $k$}) has nodes on the Fermi surface where
$\Delta$({\boldmath $k$})=0, is a particularly important theme of
research~\cite{thal}.  Many analyses have focused on the scenario
of superconductivity (SC) mediated by low energy magnetic
fluctuations, often in proximity to a quantum critical point
(QCP), where magnetic ordering temperature is driven to zero by an
external parameter such as pressure or chemical substitution.
Indeed, unconventional SC appears in the vicinity
of an antiferromagnetic (AF) QCP in most Ce-based HF compounds,
including CeIn$_3$, CePd$_2$Si$_2$ \cite{math98},
CeCoIn$_5$ \cite{pet01} and CeRhIn$_5$~\cite{heg00}, as well as organic and high-$T_c$ superconductors.

Notable counter examples have been recently reported in two Ce
compounds, in which two distinct domes of different HF
superconducting phases appear as a function of pressure or
chemical substitution.  The first example is the prototypical
 CeCu$_2$Si$_2$, with one SC dome at low pressure around
the AF QCP, and another dome emerging at high pressure distant
from the QCP~\cite{yua03}.   The superconductivity in the low
pressure dome is consistent with the magnetically mediated
pairing. On the other hand, Cooper paring resulting from the
Ce-valence fluctuations was proposed for the high pressure region,
where no discernible AF fluctuations are found~\cite{yua03,hol07}.

The second example is a quasi-two dimensional (2D)
CeIrIn$_5$($T_c$=0.4~K) \cite{pet01}, whose phase diagram is shown
in the inset of Fig.~1 \cite{nic04,kaw06}. In this system the Rh
substitution for Ir increases the $c/a$ ratio, acting as a
negative chemical pressure that increases AF correlations.  In
CeRh$_{1-x}$Ir$_x$In$_5$, the ground state continuously evolves
from AF metal (AFM) ($x<0.5$) to superconductivity ($x>0.5$).
$T_c$ shows a maximum at $x\sim 0.7$ and exhibits a cusp-like
minimum at $x \sim$0.9, forming a first dome (SC1). The strong
AF fluctuations associated with the AF QCP nearby are observed in
SC1 \cite{kaw06,kaw05}.  In CeIrIn$_5$   ($x$=1), $T_c$ increases
with pressure and exhibits a maximum ($T_c$=1~K) at $P\sim$3~GPa,
forming a second dome (SC2).  The AF fluctuations are strongly
suppressed with pressure, in SC2, far from the AF
QCP \cite{kaw06,kaw05,zhe01}. Moreover, the nature of the AF
fluctuations in magnetic fields in CeIrIn$_5$ is very different
from that in CeCoIn$_5$ and  AF CeRhIn$_5$
\cite{par06,cap04,nai07}.  Thus the superconductivity in
CeIrIn$_5$ at ambient pressure may be distinct from that in
CeCoIn$_5$ and CeRhIn$_5$, although all three compounds share
similar quasi-2D band structure~\cite{set07}.

Hence a major outstanding question is the nature of the
microscopic pairing interaction responsible for the
superconductivity in CeIrIn$_5$.  By analogy with CeCu$_2$Si$_2$,
CeIrIn$_5$ in SC2 was suggested to be a strong candidate for the
Ce-valence fluctuation mediated superconductor \cite{hol07}. To
elucidate the pairing interaction, the identification of the
superconducting gap structure is of primary importance.
Measurements of nuclear quadrupole resonance relaxation
rate~\cite{zhe01,kaw05}, thermal conductivity~\cite{mov01,tan07}
and heat capacity~\cite{mov01} revealed that the superconductivity
of CeIrIn$_5$ is unconventional,  with line nodes in the gap. Very
recently, from the measurements of the anisotropy between the
interplane and intraplane thermal conductivity $\kappa$, the gap
function of CeIrIn$_5$ was suggested to be of hybrid type,
$k_z(k_x+ik_y)$, or $E_g$ symmetry \cite{tan07}, similar to
UPt$_3$ \cite{BLussier:1994}, and  in sharp contrast to the
$d_{x^2-y^2}$ gap in CeCoIn$_5$ (and most likely in CeRhIn$_5$
under pressure) \cite{mat06}. However, as pointed out in Ref.~\cite{vek07a}, the anisotropy of $\kappa$ alone is not
sufficient to establish the hybrid gap, and further experiments
aimed at clarifying the shape of the superconducting gap are
strongly required.

In this Letter, to shed light on the pairing mechanism  of
CeIrIn$_5$, we performed the thermal transport measurements in
magnetic fields {\boldmath $H$} rotated relative to the crystal
axes.  We provide strong evidence that the gap symmetry is
$d_{x^2-y^2}$ of $B_{1g}$ symmetry.  These results put a
constraint on the pairing mechanism in CeIrIn$_5$.

Single crystals were grown by the self-flux method. The bulk
transition temperature is 0.4~K, and upper critical fields
parallel to the $ab$-plane and the $c$-axis,  $H_{c2}^{ab}$ and
$H_{c2}^{c}$,  are 1.0~T and 0.5~T at $T$=0~K, respectively.    We
measured the thermal conductivity $\kappa$ along the tetragonal
$a$-axis (heat current {\boldmath $q$}$\parallel a$) on the sample
with a rectangular shape ($2.8\times 0.45\times 0.10$~mm$^3$) by
the standard steady state method.   To apply {\boldmath $H$} with
high accuracy (misalignment of less than 0.05$^{\circ}$) relative
to the crystal axes, we used a system with two superconducting
magnets generating {\boldmath $H$} in two mutually orthogonal
directions and dilution refrigerator equipped on a mechanical
rotating stage at the top of the Dewar.

\begin{figure}[t]
\begin{center}
\includegraphics[width=7.5cm]{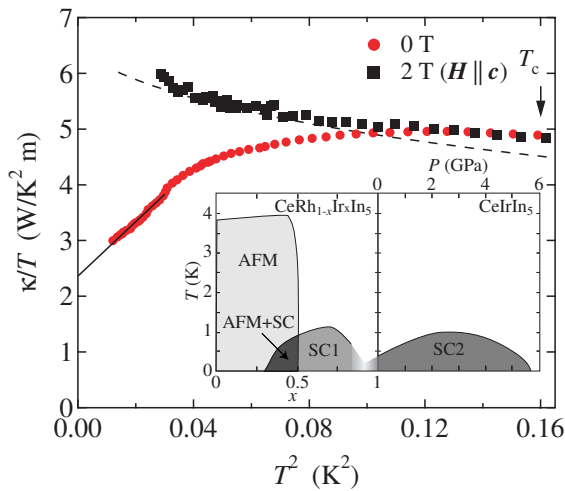}
\caption{(Color online) Temperature dependence of $\kappa/T$ as a
function of $T^2$  in zero field and above $H_{c2}^c$ ($H$=2~T).
Solid line represents $\kappa/T=\kappa_{00}/T +AT^2$; dashed line
shows $\kappa/T$ obtained from the resistivity using
Wiedemann-Franz law.  Inset: phase diagram of CeIrIn$_5$ as a
function of Ir concentration and pressure, from
Refs.\cite{nic04,kaw06}.}
\end{center}
\end{figure}

Figure~1 depicts the temperature dependence of $\kappa/T$ as  a
function of $T^2$ at $H=0$ and in the normal state above
$H_{c2}^c$. The overall behavior of  $\kappa/T$ is similar to
those reported in Refs.~\cite{mov01} and \cite{tan07}.   In zero
field, $\kappa/T$ decreases with decreasing $T$ after showing a
broad maximum at $T_c$,  similar to UPt$_3$~\cite{sud98} and
CePt$_3$Si~\cite{izaSi}.   The value of $\kappa/T$ at $T_c$ is
nearly 30~\% smaller than that reported in Ref.~\cite{tan07} and
nearly three times larger than that reported in Ref.~\cite{mov01}.
The dashed line is $\kappa/T$ obtained from the Wiedemann-Franz
law, $\kappa/T=L_0/\rho$, where
$L_0=\frac{\pi^2}{3}\left(\frac{k_B}{e}\right)^2$ is the
Sommerfeld value and $\rho$ is the measured normal state
resistivity. The observed $\kappa/T$ is close to $L_0/\rho$, which
indicates that the heat transport is dominated by the electronic
contribution.

We first discuss the thermal conductivity in zero field.  As shown
in Fig.~1, $\kappa/T$  at low $T$ is well fitted by
$\kappa/T=\kappa_{00}/T+AT^2$.  The presence of a residual term,
$\kappa_{00}/T$, as $T\rightarrow 0$~K is clearly resolved.  This
term indicates the existence of a residual normal fluid, which is
expected for nodal superconductors with impurities. For a gap with
line nodes, the magnitude of $\kappa_{00}/T$ is independent of
impurity concentration \cite{tai97}.  The present value of
$\kappa_{00}/T\simeq 2.2$~W/K$^2$m is close to that reported in
Ref.~\cite{tan07}, and,  as discussed there, is in good agreement
with the theoretical estimate for a superconductor with line node.
Thus our thermal conductivity results are also consistent with the
presence of line node in the gap function.

\begin{figure}[t]
\begin{center}
\includegraphics[width=8.5cm]{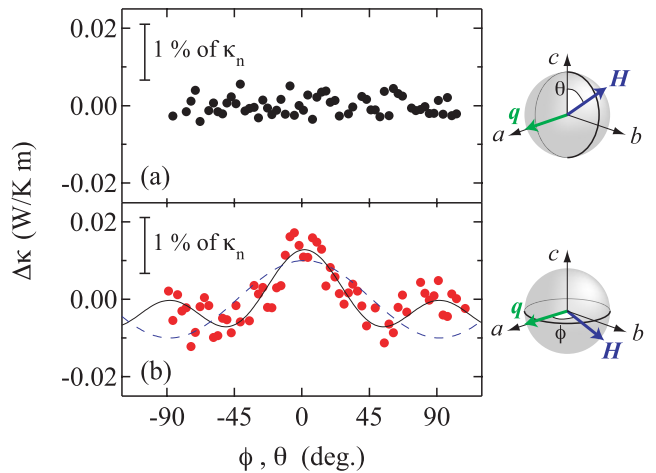}
\caption{(Color online) Angular variation of the thermal
conductivity,   $\Delta \kappa \equiv \kappa-\kappa_{0}$, at
$T$=200~mK with rotating {\boldmath $H$} (a) within the $bc$-plane
as a function of polar angle $\theta$ ($|${\boldmath
$H$}$|$=0.05~T) and (b) within the 2D $ab$-plane as a function of
azimuthal angle $\phi$ ($|${\boldmath $H$}$|$=0.10~T).  Here
$\kappa_{0}$ is the angular average of $\kappa$.  The thermal
current {\boldmath $q$} is applied parallel to the $a$-axis.  The
dashed line shows the twofold ($\cos 2\phi$) variation. The solid
line is the fit by $\kappa(\phi)= \kappa_0+C_{2 \phi}\cos
2\phi+C_{4 \phi}\cos 4\phi$, where $\kappa_0,C_{2\phi}$, and
$C_{4\phi}$ are constants. }
\end{center}
\end{figure}

The next important question is the nodal topology.   Thermal
conductivity is a powerful directional probe of the nodal
structure: Recent measurements of both $\kappa$ and the heat
capacity with {\boldmath $H$} applied at varying orientation
relative to the crystal axes established the superconducting gap
structure in {\boldmath $k$}-space in several
systems~\cite{mat06,TPark:2004,sak07}. In contrast to fully gapped
superconductors, the heat transport in nodal superconductors is
dominated by delocalized near-nodal quasiparticles. Applied field
creates a circulating supercurrent flow $\bm v_s(\bm r)$
associated with vortices. The Doppler shift of the energy of a
quasiparticle with momentum {\boldmath $p$}, $[$$E$({\boldmath
$p$})$\rightarrow E$({\boldmath $p$})--{\boldmath
$v$}$_s\cdot${\boldmath $p$}$]$, is important near the nodes,
where the local energy gap is small, $\Delta(\widehat{\bm
p})<|${\boldmath $v$}$_s\cdot${\boldmath $p$}$|$. Consequently the
density of states (DOS) depends sensitively on the angle between
{\boldmath $H$} and nodal directions \cite{IVekhter:1999}. Clear
twofold or fourfold oscillations of thermal conductivity and heat
capacity associated with the nodes have been observed in
UPd$_2$Al$_3$~\cite{wat04}, YBa$_2$Cu$_3$O$_7$~\cite{beh},
CeCoIn$_5$~\cite{sak07,izaCe} and
$\kappa$-(BEDT-TTF)$_2$Cu(NCS)$_2$~\cite{izaBE} when {\boldmath
$H$} is rotated relative to the crystal axes.

Figures~2(a) and (b) show the angular variation of the thermal
conductivity at 200~mK ($k_BT/\Delta\sim$ 0.2) as {\boldmath $H$}
is rotated within the $bc$-plane at $|${\boldmath $H$}$|$=0.05~T
($H/H_{c2}^c(T)\simeq$0.14) and within the 2D $ab$-plane at
$|${\boldmath $H$}$|$=0.1~T ($H/H_{c2}^{ab}(T) \simeq$0.14),
respectively.  Here $\theta=$({\boldmath $H$},$c$) and
$\phi=$({\boldmath $H$},$a$)  are the polar and azimuthal angles,
respectively.  For the field rotated within the $ab$-plane
($\theta=90^\circ$) $\kappa(\phi$) exhibits a distinct oscillation
as a function of $\phi$, which is characterized by peaks at
$\phi=0^{\circ}$ and $\pm90^{\circ}$ and minima at around
$\pm45^{\circ}$.    As shown by the solid line, $\kappa(\phi$) can
be decomposed into three terms,
$\kappa(\phi)=\kappa_0+\kappa_{2\phi}+\kappa_{4\phi}$, where
$\kappa_{2\phi}=C_{2 \phi}\cos 2\phi$ and $\kappa_{4\phi}=C_{4
\phi}\cos 4\phi$ have the two and four fold symmetry with respect
to $\phi$, respectively.   We note that, as shown by the dashed
line, $\kappa(\phi)$ with minima at $\pm45^{\circ}$ and peaks at
$\pm90^{\circ}$ cannot be fitted only by $\kappa_{2\phi}$-term,
indicating the presence of the fourfold term.  In sharp contrast
to {\boldmath $H$} rotating within the $ab$-plane,  no oscillation
is observed when rotating {\boldmath $H$} within the $bc$-plane;
the amplitude of the oscillation is less than 0.2\% of $\kappa_n$
if it exists, where $\kappa_n$ is the normal state thermal
conductivity measured above $H_{c2}$.  The $\kappa_{2 \phi}$ term arises from the
difference between transport parallel and perpendicular to the
vortices. Since for {\boldmath $H$} within the $bc$-plane the
field is always normal to {\boldmath $q$}, $\kappa_{2 \phi}$ term
is absent for this geometry.

\begin{figure}[t]
\begin{center}
\includegraphics[width=9.5cm]{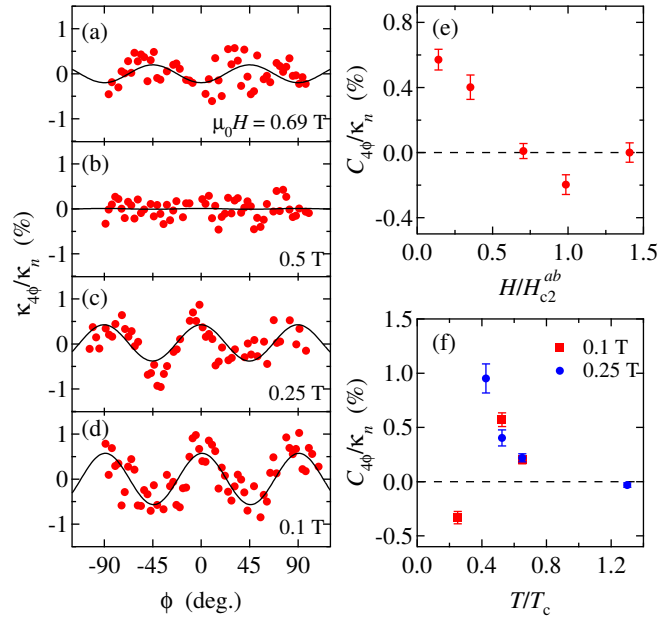}
\caption{(Color online) (a)-(d) The fourfold component,
$\kappa_{4\phi}$, normalized by $\kappa_n$ with
{\boldmath $H$} rotated in the $ab$ plane at $T$=200~mK at
$|${\boldmath $H$}$|$=0.69, 0.5, 0.25, and 0.1~T, respectively.
The upper critical field is $H_{c2}^{ab}\agt$0.7~T at this
temperature. The solid lines represent the fit by $C_{4\phi}\cos
4\phi$.  (e) $C_{4\phi}/\kappa_n$ at $T$=200~mK  as a function of
$H/H_{c2}^{ab}$.  (f) $C_{4\phi}/\kappa_n$ at $H$=0.1 and 0.25~T
as a function of $T/T_c$.  }
\end{center}
\end{figure}

We address the origin of the fourfold oscillation.   Figures
3(a)-(d) display $\kappa_{4\phi}$ normalized by $\kappa_n$ at
200~mK after the subtraction of $\kappa_0$ and $\kappa_{2\phi}$
below $H_{c2}^{ab}(T)$ ($\simeq0.7$~T). In the normal state above
$H_{c2}^{ab}$, no discernible oscillation was observed (not
shown).    At 0.69~T just below $H_{c2}^{ab}(T)$, $\kappa_{4\phi}$
exhibits a minimum at $\phi=0^{\circ}$ ($C_{4\phi}<0)$.  At
$H$=0.5~T, $\kappa_{4\phi}$ oscillation vanishes.   Further
decrease of $H$ leads to the appearance of distinct
$\kappa_{4\phi}$ oscillation that exhibits maximum at
$\phi=0^{\circ}$ ($C_{4\phi}>0$) at $H$=0.1 and 0.25~T.   Figure
3(e) shows the $H$-dependence of $C_{4\phi}/\kappa_n$.  There are
two possible origins for the fourfold oscillation: (i) the nodal
structure and (ii) in-plane anisotropy of the Fermi surface and
$H_{c2}^{ab}$.   It should be stressed that the sign of
$C_{4\phi}$ just below $H_{c2}^{ab}$ is the same as that expected
from the in-plane anisotropy of $H_{c2}^{ab}$ ($H_{c2}\parallel
(100)>H_{c2}\parallel (110)$)~\cite{wei06}, whereas its sign at low
fields is opposite.  This immediately indicates that the origin of
the fourfold symmetry at low fields is not due to the anisotropy
of the Fermi surface or $H_{c2}^{ab}$.  Rough estimate of the
amplitude of the fourfold term in layered $d$-wave superconductors
yields $C_{4\phi}/\kappa_n=0.082\frac{v_Fv'_{F}eH}{3\pi \Gamma
\Delta} \ln(\sqrt{32 \Delta / \pi \hbar \Gamma})$, where $\Delta$
is the superconducting gap, $\Gamma$ is the QP relaxation rate,
$v_F$ and $v'_F$ are the in-plane and out-of-plane Fermi
velocities \cite{won}.  Using $\Gamma\sim1.3\times10^{11}$~s$^{-1},
2\Delta/k_BT_c\sim 5, v_F\sim 1\times10^4$~m/s, and  $v'_F\sim
5\times 10^3$~m/s \cite{kaw05,izaCe} gives $C_{4\phi}/\kappa_n\sim
$2\%, of the same order as the data.  These results lead us to
conclude that  that the fourfold symmetry at low fields originates
from the nodal structure.

\begin{figure}[t]
\begin{center}
\includegraphics[width=7cm]{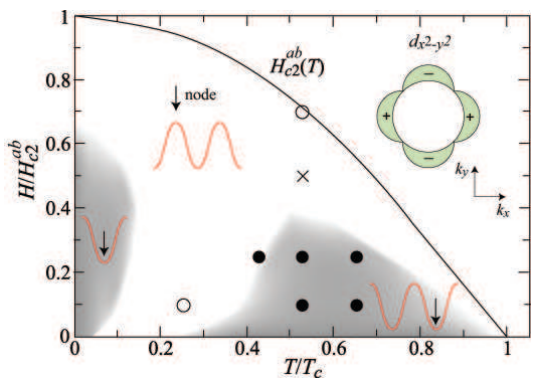}
\caption{(Color online) $H-T$ phase diagram for the fourfold term,
$\kappa_{4\phi}$, in {\boldmath $H$} rotated within the 2D
$ab$-plane.  The solid (open) circles indicate the points where
$\kappa_{4\phi}$ shows a maximum (minimum) at $\phi=0^{\circ}$.
The shaded regions correspond to the region where $\kappa_{4\phi}$
has a minimum at the nodal direction, while $\kappa_{4\phi}$ has a
maximum at the nodal directions outside this region \cite{vek07b}.
}
\end{center}
\end{figure}

The distinct fourfold oscillation within the  $ab$-plane, together
with the absence of the oscillation within the $bc$-plane,
definitely indicates the vertical line nodes perpendicular to the
$ab$-plane, and excludes a horizontal line of nodes at least in 
the dominant heavy electron bands.  Recall that
in UPd$_2$Al$_3$ with horizontal line node, clear oscillations of
$\kappa(\theta)$ are observed when rotating {\boldmath $H$} within
the $ac$-plane~\cite{wat04}. One could argue that the absence of
oscillations within the $bc$-plane shown in Fig.~2(a) is due to
relatively high temperature ($k_BT/\Delta\sim$0.2), but the
simultaneous observation of the fourfold oscillation within the
$ab$-plane at the same temperature rules out such a possibility.

Thus the superconducting symmetry of CeIrIn$_5$ is narrowed  down
to either $d_{x^2-y^2}$ or $d_{xy}$. Further identification relies
on the evolution of the oscillations with temperature and field.
In the low-$T$, low-$H$ limit, the Doppler shifted DOS shows a
maximum (minimum) when {\boldmath $H$} is along the antinodal
(nodal) directions. However, according to recent microscopic
calculations, the pattern is inverted at higher $T,H$ due to
vortex scattering, and the fourfold components of the specific
heat and of the thermal conductivity have similar behavior across
the phase diagram~\cite{vek06,vek07b}. In Fig.~3(f) we plot
$C_{4\phi}/\kappa_n$ as a function of temperature.  At $H$=0.1~T,
the sign change indeed occurs at $T/T_c\simeq 0.25$. Figure~4
displays $H-T$ phase diagram for the fourfold component. The solid
(open) circles represent the points at which observed
$\kappa_{4\phi}$ exhibits a maximum (minimum) at $\phi=0^{\circ}$,
and the shading indicates the calculated anisotropy of the thermal
conductivity for a $d$-wave superconductor from Ref.~\cite{vek07b}.
The shaded (unshaded) regions correspond to the region where
$\kappa_{4\phi}$ has a minimum (maximum) for the field in the
nodal direction. The calculation was done for a corrugated
cylindrical Fermi surface, similar to that of the main FS sheet of
CeIrIn$_5$, and the results well reproduce the observed sign
change of $\kappa_{4\phi}$. Since the minimum (maximum) of
$\kappa_{4\phi}$ occurs at $\phi=45^{\circ}$ inside (outside) the
shaded region, the nodes are located at $\pm 45^{\circ}$.  We thus
pin down the gap symmetry of CeIrIn$_5$ as $d_{x^2-y^2}$. While
reconciling the existence of vertical lines of nodes with the
results of Ref.~\cite{tan07} requires deviations from the perfect
cylindrical symmetry of the Fermi surface ~\cite{vek07a}, in the
absence of detailed calculations for the realistic band structure
of CeIrIn$_5$ (which would be desirable), the agreement with the
computed phase diagram is remarkably good.

The $d_{x^2-y^2}$ symmetry implies that the superconductivity  is
most likely to be mediated by the AF spin fluctuations, not by the
Ce-valence fluctuations \cite{tanaka07}.  Our result is also at odds with the
hybrid gap function with horizontal node proposed in
Ref.~\cite{tan07}. To our knowledge, CeIrIn$_5$ is the first
Ce-based HF compound in which $d_{x^2-y^2}$ symmetry is realized
in a distinct superconducting phase remote from the AF QCP.   It
is intriguing that the superconductivity in SC1 and SC2 phases has
the same gap symmetry.  A possible explanation for the double dome structure is that
the active area on the Fermi surface for the superconductivity,
which is nested by magnetic propagation vector {\boldmath $Q$}=$(\frac{1}{2}, \frac{1}{2}, Q_z)$, has different $Q_z$ in
these two phases.  In fact, a remarkable difference between
CeCoIn$_5$ and CeIrIn$_5$ is that the incommensurate AF order with $Q_z$=0.298
strongly suppresses the superconductivity in
CeRh$_{1-x}$Co$_x$In$_5$ \cite{seiko07}, while they coexist in
CeRh$_{1-x}$Ir$_x$In$_5$ \cite{llo05}.

In conclusion, the measurements of the thermal conductivity  under
rotated magnetic fields provide a strong evidence that the
superconducting gap of CeIrIn$_5$ at ambient pressure has vertical
line nodes, and is of $d_{x^2-y^2}$ symmetry.  This indicates that
two distinct domes of HF superconducting phases possess the same
superconducting symmetry, in which AF fluctuations appear to play
an important role.  The determined gap symmetry in CeIrIn$_5$ remote from the
AF QCP further restricts theories of the pairing mechanism.

We thank H.~Ikeda, K.~Kontani, P.~Thalmeier, and S.~Watanabe for
helpful discussions.

\end{document}